\documentclass[aps,prl,showpacs,twocolumn,amsmath,amssymb,10pt]{revtex4-1}
\usepackage{graphicx}
\usepackage{siunitx}
\usepackage{color}

\newcommand{\pd}[2]{\frac{\partial #1}{\partial #2}} 
\newcommand{\mean}[1]{\left\langle #1 \right\rangle} 

\begin{document}

\title{Noisy Nonlinear Dynamics of Vesicles in Flow}
\author{David Abreu and Udo Seifert}
\affiliation{{II.} Institut f\"ur Theoretische Physik, Universit\"at Stuttgart, 70550 Stuttgart, Germany}
\date{\today}
\pacs{87.16.D-, 47.63.-b, 05.40.-a}

\begin{abstract}
We present a model for the dynamics of fluid vesicles in linear flow which consistently includes thermal fluctuations and nonlinear coupling between different modes. At the transition between tank treading and tumbling, we predict a trembling motion which is at odds with the known deterministic motions and for which thermal noise is strongly amplified. In particular, highly asymmetric shapes are observed even though the deterministic flow only allows for axisymmetric ones. Our results explain quantitatively recent experimental observations [Levant and Steinberg, Phys. Rev. Lett. 109, 268103 (2012)].
\end{abstract}

\maketitle

{\sl Introduction.---} 
For a system close to a dynamical instability, the effect of small perturbations can be dramatic \cite{grah74,wies85,krav03}. In particular, external noise \cite{wies85} such as thermal fluctuations may be strongly amplified in the vicinity of a bifurcation, therefore crucially affecting the dynamics of the system. Such effects were observed, e. g., in binary fluids \cite{schoe92}, fiber lasers \cite{huer09}, and even population dynamics \cite{reum06}. Recently, Levant and Steinberg \cite{leva12a} realized experiments in which such an amplification of thermal noise occurred for fluid vesicles subject to an external flow. 

Fluid vesicles \cite{seif97} are closed lipid bilayer membranes which mimic biological objects such as red blood cells. Understanding their dynamics in flow may therefore help to elucidate blood rheology \cite{abka08,fink11,fedo11,misb12,dupi12} or chemical activity \cite{fors11}. This dynamics is strongly nonlinear due to curvature elasticity and the incompressibility of the fluid membrane. Most theoretical models \cite{vlah07,lebe07,lebe08,faru10,faru12a,faru12b} and numerical simulations \cite{bibe11,zhao11,sala12,yazd12} neglect thermal fluctuations, mainly because the bending rigidity of the membrane is of the order of $\kappa\simeq 25k_BT$. For weak flows obeying Stokes equation, these models predict three main dynamical regimes: (i) a tank-treading (TT) motion, for which the orientation of the vesicle remains constant while the membrane rotates around the interior fluid; (ii) a solidlike rotation of the vesicle called tumbling (TB); (iii) and a vacillating-breathing (VB) motion \cite{misb06,gued12} (also called swinging \cite{nogu07} or trembling \cite{lebe07}) 
at the TT-TB transition, in which the orientation angle oscillates around $0$ whereas the shape periodically deforms. An important feature of these predictions is that the vesicle shape remains axisymmetric at all times.

The TT and TB motion observed in experiments \cite{kant05,kant06,made06} are well described by the theoretical predictions. However, a series of experiments \cite{kant06,desc09,desc09a,zabu11,leva12a} shows that the trembling (TR) motion found at the TT-TB transition is very different from the deterministic VB motion. The experimental TR is an aperiodic motion characterized by highly asymmetric shapes and a larger basin of attraction in the parameter space than VB. The discrepancy between models and experiments is presumably due to thermal fluctuations \cite{zabu11,leva12a}, a claim supported by some qualitative results obtained in stochastic simulations \cite{nogu04,nogu05,me09} which, however, have not looked systematically at the TT-TB transition. Analytical models including thermal fluctuations \cite{seif99,fink08,abre12} exist but have not yet addressed the TR motion. 

In this Letter, we present a model for vesicles in planar linear flow which systematically and consistently includes thermal fluctuations. We predict a trembling regime at the TT-TB transition which is at odds with deterministic predictions but agrees very well with the experimental observations. We show that, because of the amplification of thermal noise, highly asymmetric shapes are observed even though the deterministic external flow would only allow for axisymmetric ones.

{\sl Model.---} 
We consider a vesicle of volume $V\equiv4\pi R_0^3/3$ and surface area $A\equiv(4\pi+\Delta)R_0^2$, where $R_0$ is the radius of a sphere of the same volume and $\Delta$ the excess area relative to it. This vesicle encloses a fluid of viscosity $\eta_i$ and is suspended in another liquid of viscosity $\eta_o$. The viscosity contrast $\lambda\equiv\eta_i/\eta_o$ is set to $1$ in the following. The two-dimensional fluid membrane is assumed to be incompressible and impermeable so that $V$ and $A$, or equivalently $R_0$ and $\Delta$, remain constant. This vesicle is subject to a planar linear flow, for which the unperturbed velocity field can be written as 
  \begin{equation}
    {\bf v}=s(x{\bf e_y}+y{\bf e_x})+\omega(y{\bf e_x}-x{\bf e_y})
    \label{flow}
  \end{equation}
where $s$ is the strength of the extensional flow, which tends to stretch the vesicle along the ($y$=$x$)-direction, and $\omega$ the vorticity, which is the strength of the rotational component of the flow. A pure shear flow corresponds to $s=\omega$. We consider weak flows such that the Reynolds number is low and the hydrodynamics is described by the Stokes equation. Moreover, the energetic cost of the vesicle deformation is given by the Helfrich energy \cite{seif97}
  \begin{equation}
    {\cal H}=\int dA\left[\frac{\kappa}{2}(2H)^2+\sigma\right]
    \label{helf}
  \end{equation}
where $\kappa$ is the bending rigidity of the membrane, $H$ the local mean curvature, and $\sigma$ the surface tension ensuring local and global incompressibility.

In order to derive analytically the equations of motion, we assume that the vesicle is quasispherical; i. e., its excess area $\Delta$ is small. We then expand the radius
  \begin{equation}
    {\bf R}(\theta,\phi)=R_0\left(1+\sum_{l=0}^{\infty}\sum_{m=-l}^l u_{lm}{\cal Y}_{lm}(\theta,\phi)\right){\bf e_r}
    \label{radius}
  \end{equation}
in the basis of spherical harmonics ${\cal Y}_{lm}$. Adapting \cite{seif99}, we solve the hydrodynamic problem inside and outside the vesicle and match the solutions at the sphere of radius $R_0$, which leads to the equations of motion
  \begin{equation}
    \dot u_{lm}=im\omega u_{lm}-\mu_l\pd{{\cal H}}{u_{lm}^*}-ihs\delta_{l,2}(\delta_{m,2}-\delta_{m,-2})+\zeta_{lm}(t)
    \label{eqmo}
  \end{equation}
for the coefficients $u_{lm}$, where the dot stands for the time derivative and $h\equiv24\sqrt{2\pi/15}/11$, $\mu_l\equiv\Gamma_l/\eta_oR_0^3$ and $\Gamma_l\equiv l(l+1)/(4l^3+6l^2-1)$. The thermal noise $\zeta_{lm}(t)$ is white and its correlations
  \begin{equation}
    \mean{\zeta_{lm}(t)\zeta_{l'm'}(t')}=2\mu_lk_BT(-1)^m\delta_{l,l'}\delta_{m,-m'}\delta(t-t')
    \label{noise}
  \end{equation}
are given by the fluctuation-dissipation theorem in equilibrium. The first term on the right-hand side of Eq. \eqref{eqmo} is a norm-preserving rotation of $u_{lm}$ for all harmonics and is due to the vorticity $\omega$ of the external flow. The third term is only present for the prolate harmonics ${\cal Y}_{2,\pm2}$, which are the only ones directly excited by the flow. For this reason, we can distinguish between different regimes by looking at the evolution of the orientation angle $\phi_{22}$ of the prolate harmonic defined by $u_{22}\equiv |u_{22}|\mathrm{e}^{-2i\phi_{22}}$. It corresponds to the inclination angle of the long axis of the vesicle in the flow plane. Its equation of motion follows from \eqref{eqmo} as
  \begin{equation}	
    \dot\phi_{22}=-\omega\left(1-\frac{h\cos(2\phi_{22})}{2|u_{22}|}\frac{s}{\omega}\right)+\frac{\zeta_\phi(t)}{2\sqrt{2}|u_{22}|}
    \label{angle}
  \end{equation}
where $\zeta_\phi$ is a Gaussian white noise having the correlations \eqref{noise}. We see that for small values of the ratio $\omega/s$, the deterministic part of Eq. \eqref{angle} may have a fixed point, whereas for larger vorticities, the right-hand side is always negative. These two cases correspond roughly to the TT and TB motions, respectively. 

The nonlinearity of the equations \eqref{eqmo} is due to the term depending on the Helfrich energy \eqref{helf}. Following Ref. \cite{lebe07}, we expand the bending energy ${\cal H}$ to third order in the coefficients $u_{lm}$ \cite{zho89}. Such terms are necessary for the dynamics to be insensitive to initial conditions \cite{lebe08}. We do not include the hydrodynamic corrections discussed in Refs. \cite{dank07,faru10}, but keep in mind that they become relevant for strong enough flows. Through the condition of constant volume, we can express $u_{00}$ as a function of all other coefficients with $l\geq1$. The homogeneous part of the surface tension $\sigma$ has to be determined such that $\Delta$ remains constant. The local incompressibility of the membrane has already been taken into account in deriving Eq. \eqref{eqmo}. 

\begin{figure}
  \centering
  \includegraphics[width=0.99\linewidth]{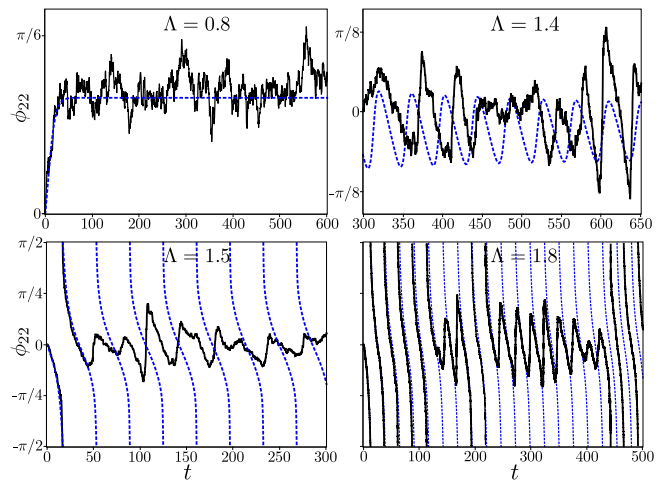}
  \caption{Temporal evolution of the inclination angle $\phi_{22}$ (solid black curves) for $\Delta=0.3$, $S=30$ ($s=\SI{0.032}{\second^{-1}}$) and $\Lambda=0.8$, $1.4$, $1.5$, and $1.8$  ($\omega=\SI{0.066}{\second^{-1}}$, $\SI{0.11}{\second^{-1}}$, $\SI{0.12}{\second^{-1}}$ and $\SI{0.15}{\second^{-1}}$).  The blue dashed curves are the corresponding deterministic trajectories.}
  \label{4reg}
\end{figure}

In the deterministic case, only the harmonics of order $l=0$, $2$, and $4$ survive to that order of expansion \cite{faru10}. In particular, terms with odd $l$ remain strictly equal to zero. Now, if we include thermal fluctuations, all $u_{lm}$ have a positive mean absolute value. Therefore, they cannot be discarded out of symmetry considerations since they might be important at the dynamical transitions. We will thus keep all harmonics between $l=2$ and $l=5$ (the modes with $l=1$ correspond to a translation of the vesicle as a whole and are not relevant to our problem) and solve numerically the eighteen coupled equations of motion. In the numeric scheme, we replace the surface tension $\sigma$, which strictly enforces the constraint of constant area, by a softer energetic constraint of the form $K(A^{(3)}-A)^2/2$ in the bending energy \eqref{helf}, where $A^{(3)}$ is the expansion of the area $A$ to third order in $u_{lm}$. For $K\to\infty$, we thus recover the same dynamics as with the rigid constraint $\sigma$. The soft constraint, however, is much easier to treat numerically since it avoids the intricacies of multiplicative noise \cite{mors04}. Furthermore, in order to compare our results to other models and to experiments, we introduce the parameters \cite{lebe07}
  \begin{align}
    S\equiv\frac{14\pi}{3\sqrt{3}}\frac{\eta_oR_0^3}{\kappa\Delta}s & &\Lambda\equiv\frac{55}{24}\sqrt{\frac{3\Delta}{10\pi}}\frac{\omega}{s},
    \label{adim}
  \end{align}
where $S$ is the dimensionless strength of the elongational flow and $\Lambda$ is the rescaled ratio of the rotational and elongational components of the flow. It has been claimed that these two quantities are enough to describe the dynamics of the system \cite{lebe07}, a fact contested in other theoretical models \cite{kaou09a} but found to be pretty accurate in experiments \cite{desc09}.  In the following, we set $T=\SI{293}{\kelvin}$, $R_0=\SI{15}{\micro\meter}$, $\kappa=25k_BT$ and $\eta_o=\SI{1}{\milli\pascal\second}$, which correspond to typical experimental values \cite{leva12a}. The comparison to the deterministic case is done by ignoring the thermal noise in our equations.

{\sl Regimes of motion.---} Figure \ref{4reg} shows the temporal evolution of the inclination angle $\phi_{22}$ for $S=30$ and four different values of $\Lambda$. For small values of the vorticity ($\Lambda=0.8$), the vesicle is in the TT regime, in which its orientation fluctuates around the deterministic value but remains positive. If we increase the vorticity of the flow, we observe the TR motion ($\Lambda=1.4$ and $1.5$), for which the oscillation angle irregularly oscillates around $0$, although some full rotation might occur from time to time (see beginning of $\Lambda=1.5$). Without thermal noise, the model predicts a periodic VB motion for $\Lambda=1.4$ but TB already for $\Lambda=1.5$, whereas the stochastic motion is still TR. This observation is in agreement with experimental results, which find that TR happens for a much broader range of parameters than predicted by deterministic 
models \cite{desc09,desc09a}. For $\Lambda=1.8$, we see a mixture between TR and TB, with bursts of TR happening between the TB cycles. Thermal fluctuations mix both regimes at the TR-TB transition, a similar effect to what happens for undeformable capsules at the transition between tumbling and swinging \cite{abre12}.

\begin{figure}
  \centering
  \includegraphics[width=0.99\linewidth]{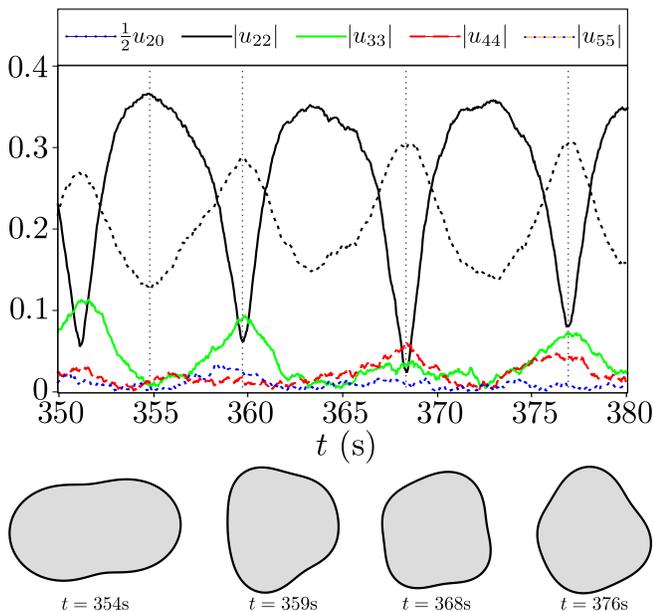}
  \caption{Top: Evolution of some mode amplitudes for $\Delta=0.66$, $S=60.7$, and $\Lambda=1.8$ (corresponding to $s=\SI{0.14}{\second^{-1}}$ and $\omega=\SI{0.44}{\second^{-1}}$). Bottom: Instantaneous shape contours in the flow plane at four different times (see vertical lines in the plot).}
  \label{trem}
\end{figure}
The difference between stochastic and deterministic models becomes even clearer if one looks at the evolution of the shape of the vesicle. Figure \ref{trem} shows the time evolution of the coefficients $u_{20}$ and $u_{ll}$ ($l=2,...,5$) for a typical mixed TR and TB motion, using the same parameters as in Fig. 2 of Ref. \cite{leva12a}. We choose these coefficients because they determine the shape of the vesicle in the flow plane, which is observed in experiments. In the deterministic case, the odd harmonics $u_{33}$ and $u_{55}$ are strictly equal to zero, whereas $u_{44}$ remains much smaller than $u_{20}$ and $u_{22}$. The shape of the vesicle remains ellipsoidal, even though there is a periodic ``breathing" due to the interplay between $u_{20}$ and $u_{22}$. With thermal noise, all modes are excited and higher harmonics become regularly larger that the second one, leading to strongly distorted shapes as shown by the four snapshots of the contour of the vesicle in the flow plane, which are similar to the experimental ones \cite{leva12a}. This shape distortion happens when the inclination angle becomes negative, i. e., when the vesicle is compressed rather than stretched by the external flow. This phenomenon is reminiscent of vesicle wrinkling \cite{kant07,fink06,turi08}, which happens when the direction of the elongational flow is suddenly inverted. However, wrinkling is a transient, deterministic phenomenon which is triggered by a discontinuity in the external flow. The trembling motion presented here exists only with thermal fluctuations and is a steady-state property of the vesicle dynamics. Note that this kind of shape deformations are also present in TB, since the vesicle also passes through negative inclination angles, but become smaller for larger $\Lambda$ because the vorticity $\omega$ and thus the rotation frequency of the vesicle become larger.

{\sl Power spectra and phase diagram.---} 
\begin{figure}
  \centering
  \includegraphics[width=0.99\linewidth]{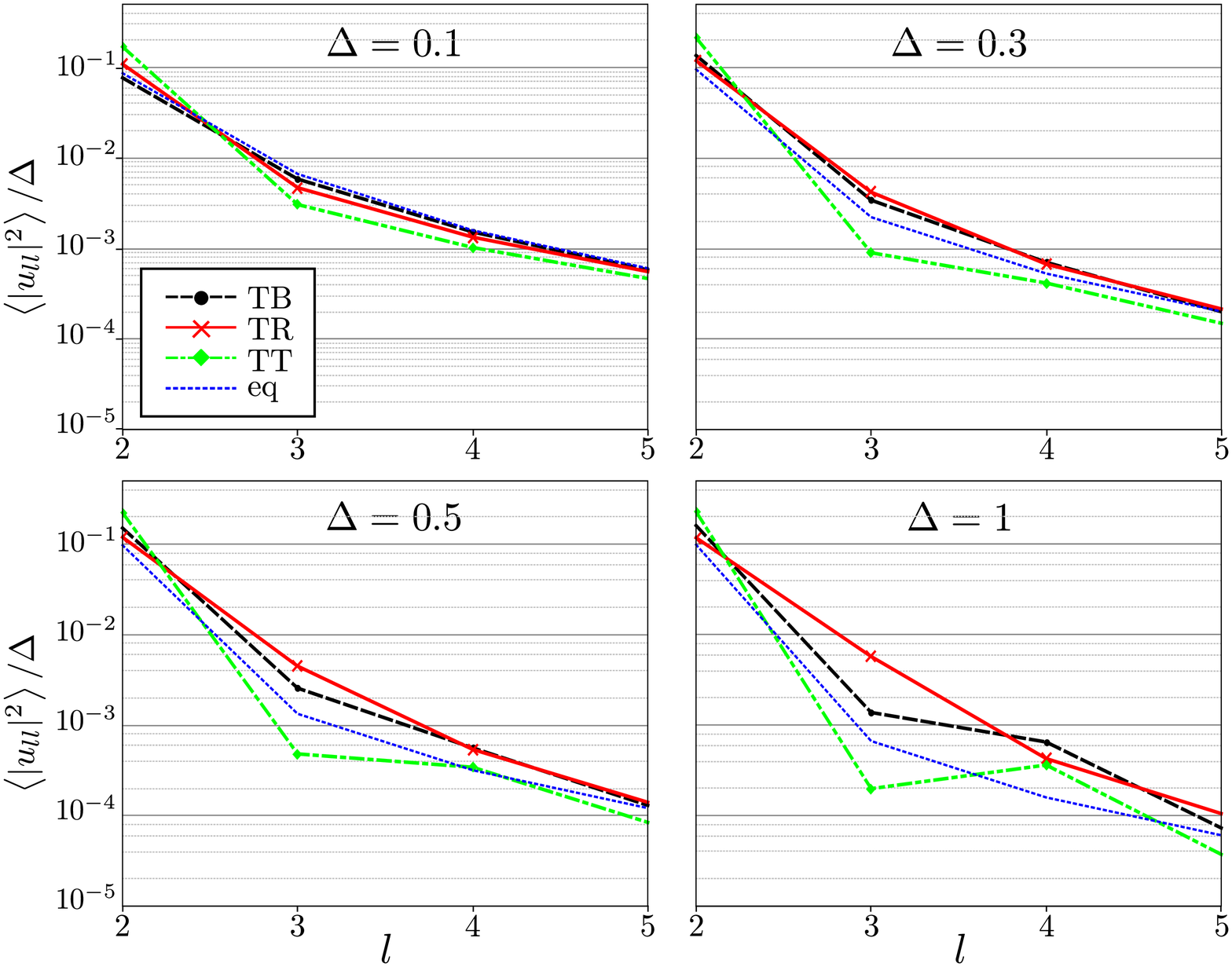}
  \caption{Power spectra of the rescaled amplitudes $|u_{ll}|^2/\Delta$ as a function of the mode number $l=2,...,5$ for $S=30$ and $\Lambda=0.8$ (TT, green dash-dotted line), $1.5$ (TR, red full line) and $2.5$ (TB, black dashed line) and four different excess areas $\Delta$. The blue dotted lines are the corresponding equilibrium spectra \cite{seif97}.}
  \label{power}
\end{figure}
We can quantify the shape dynamics by looking at the power spectra of the rescaled harmonics $|u_{ll}|^2/\Delta$ in the different regimes, as shown in Fig. \ref{power} for $S=30$ and different values of $\Delta$. For low excess areas ($\Delta=0.1$), the power spectra in TR and TB are very close to the equilibrium one due to the fact that the absolute strength $s$ of the elongational flow is relatively small ($s\simeq \SI{0.01}{\second^{-1}}$) thus exerting stresses comparable to the thermal ones. The TT spectrum already shows a larger second harmonic than in equilibrium (and correspondingly smaller higher-order harmonics) because the orientation angle $\phi_{22}$ remains positive and therefore the vesicle is constantly stretched. $S$ being constant, larger $\Delta$ means larger $s$, which should suppress the effect of thermal fluctuations. This suppression indeed happens for the TT and TB motions, for which we see a decrease of the amplitude of the odd modes while the fourth mode, which is coupled to the second one, becomes larger. By contrast, in the TR regime the third mode remains always one order of magnitude larger than the fourth one even for large $\Delta$. These features are also found in the experimental data \cite{zabu11,leva12a} although the coupling between even modes seems stronger there. The TR motion is thus directly caused by the strong amplification of thermal noise at the dynamical TT-TB bifurcation.

\begin{figure}
  \centering
  \includegraphics[width=0.99\linewidth]{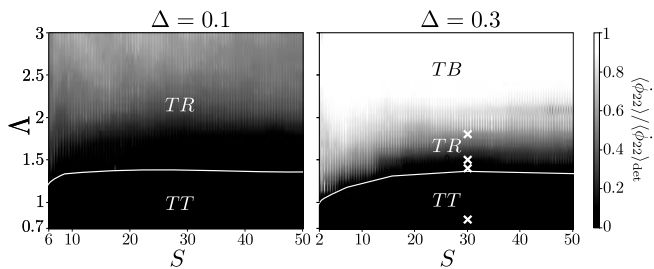}
  \caption{Mean velocity $\langle\dot\phi_{22}\rangle$ rescaled by the deterministic one $\langle\dot\phi_{22}\rangle_\mathrm{det}$ as a function of the dimensionless variables $S$ and $\Lambda$ for $\Delta=0.1$ and $0.3$. The white lines correspond to the TT-TR transition. The four crosses correspond to Fig. \ref{4reg}.}
  \label{phdiag}
\end{figure}
Further interesting features are revealed by the dynamical phase diagram of the system, as shown in Fig. \ref{phdiag} for two different excess areas. Since the transitions between the different regimes are smoothed out by thermal fluctuations, we choose the mean tumbling velocity $\langle\dot\phi_{22}\rangle$ as an order parameter. When the deterministic prediction is TB, we rescale this order parameter by the deterministic mean velocity $\langle\dot\phi_{22}\rangle_\mathrm{det}>0$ so that we clearly see the effect of thermal noise. In the TB regime, this ratio is close to $1$ (white regions) while it is almost $0$ in the TT and TR regimes (black regions). The gray regions indicate a mixed TR and TB motion (see Fig. \ref{4reg} for $\Lambda=1.8$). We further define TT as the regime in which the mean orientation angle is larger than its standard deviation, such that we are able to distinguish between TT and TR. For small enough $S$, the dynamics is dominated mainly by diffusive motion of the orientation angle, making it hard to distinguish between different regimes. For this reason, these phase diagrams do not extend to $S=0$. The first striking feature is that the TR region is much larger for $\Delta=0.1$ than for $\Delta=0.3$. This difference is due to the fact that, for fixed $S$, the magnitude of the elongation $s$ grows with $\Delta$, such that thermal fluctuations have a greater effect for small excess areas. The TR region is also much wider than the one predicted by the deterministic model of \cite{lebe07}, which corresponds approximately to our model without noise. Going to larger $\Delta$, the TR region becomes thinner, but then the neglected higher-order hydrodynamic terms become important. If we had included these, we might observe a broadening of the TR region with growing $\Delta$, similar to what happens for the VB region in deterministic models including hydrodynamic corrections \cite{faru10}. In addition, the TT-TR transition happens approximately at fixed values of $\Lambda$ for all $\Delta$. The phase diagrams for different values of $\Delta$ have therefore a large overlap, which might explain why $S$ and $\Lambda$ are so effective for classifying experimental data \cite{zabu11}. Our results show, however, that $\Delta$ is also important for the dynamics and that $S$ and $\Lambda$ are not sufficient control parameters. Finally, because of the thermal fluctuations, the size of the TR region is quantitatively comparable to the one measured in experiments \cite{desc09,desc09a,zabu11}.

{\sl Conclusion.---} 
We have shown that even for large vesicles with an effective radius of $\SI{15}{\micro\meter}$, thermal fluctuations are crucial to understand the dynamics in linear flow. This is due to the sensitivity of the nonlinear dynamics of the vesicle to small perturbations close to the TT-TB transition. In this case, the small perturbations correspond to the always present thermal fluctuations which incessantly destabilize the system, generating the trembling motion. Our results thus give an explanation for the strong shape fluctuations observed in experiments that are not predicted by deterministic models. The generalization of these results to other soft objects such as red blood cells or elastic capsules will be the focus of further investigation.

We thank V. Steinberg for discussions and acknowledge financial support from GIF.

\bibliography{/home/abreu/Documents/Literatur/doktor.bib}

\begin{thebibliography}{46}%
\makeatletter
\providecommand \@ifxundefined [1]{%
 \@ifx{#1\undefined}
}%
\providecommand \@ifnum [1]{%
 \ifnum #1\expandafter \@firstoftwo
 \else \expandafter \@secondoftwo
 \fi
}%
\providecommand \@ifx [1]{%
 \ifx #1\expandafter \@firstoftwo
 \else \expandafter \@secondoftwo
 \fi
}%
\providecommand \natexlab [1]{#1}%
\providecommand \enquote  [1]{``#1''}%
\providecommand \bibnamefont  [1]{#1}%
\providecommand \bibfnamefont [1]{#1}%
\providecommand \citenamefont [1]{#1}%
\providecommand \href@noop [0]{\@secondoftwo}%
\providecommand \href [0]{\begingroup \@sanitize@url \@href}%
\providecommand \@href[1]{\@@startlink{#1}\@@href}%
\providecommand \@@href[1]{\endgroup#1\@@endlink}%
\providecommand \@sanitize@url [0]{\catcode `\\12\catcode `\$12\catcode
  `\&12\catcode `\#12\catcode `\^12\catcode `\_12\catcode `\%12\relax}%
\providecommand \@@startlink[1]{}%
\providecommand \@@endlink[0]{}%
\providecommand \url  [0]{\begingroup\@sanitize@url \@url }%
\providecommand \@url [1]{\endgroup\@href {#1}{\urlprefix }}%
\providecommand \urlprefix  [0]{URL }%
\providecommand \Eprint [0]{\href }%
\providecommand \doibase [0]{http://dx.doi.org/}%
\providecommand \selectlanguage [0]{\@gobble}%
\providecommand \bibinfo  [0]{\@secondoftwo}%
\providecommand \bibfield  [0]{\@secondoftwo}%
\providecommand \translation [1]{[#1]}%
\providecommand \BibitemOpen [0]{}%
\providecommand \bibitemStop [0]{}%
\providecommand \bibitemNoStop [0]{.\EOS\space}%
\providecommand \EOS [0]{\spacefactor3000\relax}%
\providecommand \BibitemShut  [1]{\csname bibitem#1\endcsname}%
\let\auto@bib@innerbib\@empty
\bibitem [{\citenamefont {Graham}(1974)}]{grah74}%
  \BibitemOpen
  \bibfield  {author} {\bibinfo {author} {\bibfnamefont {R.}~\bibnamefont
  {Graham}},\ }\href {\doibase 10.1103/PhysRevA.10.1762} {\bibfield  {journal}
  {\bibinfo  {journal} {Phys. Rev. A}\ }\textbf {\bibinfo {volume} {10}},\
  \bibinfo {pages} {1762} (\bibinfo {year} {1974})}\BibitemShut {NoStop}%
\bibitem [{\citenamefont {Wiesenfeld}(1985)}]{wies85}%
  \BibitemOpen
  \bibfield  {author} {\bibinfo {author} {\bibfnamefont {K.}~\bibnamefont
  {Wiesenfeld}},\ }\href {\doibase 10.1007/BF01010430} {\bibfield  {journal}
  {\bibinfo  {journal} {J. Stat. Phys.}\ }\textbf {\bibinfo {volume} {38}},\
  \bibinfo {pages} {1071} (\bibinfo {year} {1985})}\BibitemShut {NoStop}%
\bibitem [{\citenamefont {Kravtsov}\ and\ \citenamefont
  {Surovyatkina}(2003)}]{krav03}%
  \BibitemOpen
  \bibfield  {author} {\bibinfo {author} {\bibfnamefont {Y.~A.}\ \bibnamefont
  {Kravtsov}}\ and\ \bibinfo {author} {\bibfnamefont {E.~D.}\ \bibnamefont
  {Surovyatkina}},\ }\href {\doibase 10.1016/j.physleta.2003.10.034} {\bibfield
   {journal} {\bibinfo  {journal} {Phys. Lett. A}\ }\textbf {\bibinfo {volume}
  {319}},\ \bibinfo {pages} {348 } (\bibinfo {year} {2003})}\BibitemShut
  {NoStop}%
\bibitem [{\citenamefont {Sch{\"o}pf}\ and\ \citenamefont
  {Rehberg}(1992)}]{schoe92}%
  \BibitemOpen
  \bibfield  {author} {\bibinfo {author} {\bibfnamefont {W.}~\bibnamefont
  {Sch{\"o}pf}}\ and\ \bibinfo {author} {\bibfnamefont {I.}~\bibnamefont
  {Rehberg}},\ }\href {\doibase 10.1209/0295-5075/17/4/007} {\bibfield
  {journal} {\bibinfo  {journal} {Europhys. Lett.}\ }\textbf {\bibinfo {volume}
  {17}},\ \bibinfo {pages} {321} (\bibinfo {year} {1992})}\BibitemShut
  {NoStop}%
\bibitem [{\citenamefont {Huerta-Cuellar}\ \emph {et~al.}(2009)\citenamefont
  {Huerta-Cuellar}, \citenamefont {Pisarchik}, \citenamefont {Kir'yanov},
  \citenamefont {Barmenkov},\ and\ \citenamefont {del
  Valle~Hern\'andez}}]{huer09}%
  \BibitemOpen
  \bibfield  {author} {\bibinfo {author} {\bibfnamefont {G.}~\bibnamefont
  {Huerta-Cuellar}}, \bibinfo {author} {\bibfnamefont {A.~N.}\ \bibnamefont
  {Pisarchik}}, \bibinfo {author} {\bibfnamefont {A.~V.}\ \bibnamefont
  {Kir'yanov}}, \bibinfo {author} {\bibfnamefont {Y.~O.}\ \bibnamefont
  {Barmenkov}}, \ and\ \bibinfo {author} {\bibfnamefont {J.}~\bibnamefont {del
  Valle~Hern\'andez}},\ }\href {\doibase 10.1103/PhysRevE.79.036204} {\bibfield
   {journal} {\bibinfo  {journal} {Phys. Rev. E}\ }\textbf {\bibinfo {volume}
  {79}},\ \bibinfo {pages} {036204} (\bibinfo {year} {2009})}\BibitemShut
  {NoStop}%
\bibitem [{\citenamefont {Reuman}\ \emph {et~al.}(2006)\citenamefont {Reuman},
  \citenamefont {Desharnais}, \citenamefont {Costantino}, \citenamefont
  {Ahmad},\ and\ \citenamefont {Cohen}}]{reum06}%
  \BibitemOpen
  \bibfield  {author} {\bibinfo {author} {\bibfnamefont {D.~C.}\ \bibnamefont
  {Reuman}}, \bibinfo {author} {\bibfnamefont {R.~A.}\ \bibnamefont
  {Desharnais}}, \bibinfo {author} {\bibfnamefont {R.~F.}\ \bibnamefont
  {Costantino}}, \bibinfo {author} {\bibfnamefont {O.~S.}\ \bibnamefont
  {Ahmad}}, \ and\ \bibinfo {author} {\bibfnamefont {J.~E.}\ \bibnamefont
  {Cohen}},\ }\href {\doibase 10.1073/pnas.0608571103} {\bibfield  {journal}
  {\bibinfo  {journal} {Proc. Nat. Acad. Sci. U. S. A.}\ }\textbf {\bibinfo
  {volume} {103}},\ \bibinfo {pages} {18860} (\bibinfo {year}
  {2006})}\BibitemShut {NoStop}%
\bibitem [{\citenamefont {Levant}\ and\ \citenamefont
  {Steinberg}(2012)}]{leva12a}%
  \BibitemOpen
  \bibfield  {author} {\bibinfo {author} {\bibfnamefont {M.}~\bibnamefont
  {Levant}}\ and\ \bibinfo {author} {\bibfnamefont {V.}~\bibnamefont
  {Steinberg}},\ }\href {\doibase 10.1103/PhysRevLett.109.268103} {\bibfield
  {journal} {\bibinfo  {journal} {Phys. Rev. Lett.}\ }\textbf {\bibinfo
  {volume} {109}},\ \bibinfo {pages} {268103} (\bibinfo {year}
  {2012})}\BibitemShut {NoStop}%
\bibitem [{\citenamefont {Seifert}(1997)}]{seif97}%
  \BibitemOpen
  \bibfield  {author} {\bibinfo {author} {\bibfnamefont {U.}~\bibnamefont
  {Seifert}},\ }\href {\doibase 10.1080/00018739700101488} {\bibfield
  {journal} {\bibinfo  {journal} {Adv. Phys.}\ }\textbf {\bibinfo {volume}
  {46}},\ \bibinfo {pages} {13} (\bibinfo {year} {1997})}\BibitemShut {NoStop}%
\bibitem [{\citenamefont {Abkarian}\ and\ \citenamefont
  {Viallat}(2008)}]{abka08}%
  \BibitemOpen
  \bibfield  {author} {\bibinfo {author} {\bibfnamefont {M.}~\bibnamefont
  {Abkarian}}\ and\ \bibinfo {author} {\bibfnamefont {A.}~\bibnamefont
  {Viallat}},\ }\href {\doibase 10.1039/B716612E} {\bibfield  {journal}
  {\bibinfo  {journal} {Soft Matter}\ }\textbf {\bibinfo {volume} {4}},\
  \bibinfo {pages} {653} (\bibinfo {year} {2008})}\BibitemShut {NoStop}%
\bibitem [{\citenamefont {Finken}\ \emph {et~al.}(2011)\citenamefont {Finken},
  \citenamefont {Kessler},\ and\ \citenamefont {Seifert}}]{fink11}%
  \BibitemOpen
  \bibfield  {author} {\bibinfo {author} {\bibfnamefont {R.}~\bibnamefont
  {Finken}}, \bibinfo {author} {\bibfnamefont {S.}~\bibnamefont {Kessler}}, \
  and\ \bibinfo {author} {\bibfnamefont {U.}~\bibnamefont {Seifert}},\ }\href
  {\doibase 10.1088/0953-8984/23/18/184113} {\bibfield  {journal} {\bibinfo
  {journal} {J. Phys.: Condens. Matter}\ }\textbf {\bibinfo {volume} {23}},\
  \bibinfo {pages} {184113} (\bibinfo {year} {2011})}\BibitemShut {NoStop}%
\bibitem [{\citenamefont {Fedosov}\ \emph {et~al.}(2011)\citenamefont
  {Fedosov}, \citenamefont {Pan}, \citenamefont {Caswell}, \citenamefont
  {Gompper},\ and\ \citenamefont {Karniadakis}}]{fedo11}%
  \BibitemOpen
  \bibfield  {author} {\bibinfo {author} {\bibfnamefont {D.~A.}\ \bibnamefont
  {Fedosov}}, \bibinfo {author} {\bibfnamefont {W.}~\bibnamefont {Pan}},
  \bibinfo {author} {\bibfnamefont {B.}~\bibnamefont {Caswell}}, \bibinfo
  {author} {\bibfnamefont {G.}~\bibnamefont {Gompper}}, \ and\ \bibinfo
  {author} {\bibfnamefont {G.~E.}\ \bibnamefont {Karniadakis}},\ }\href
  {\doibase 10.1073/pnas.1101210108} {\bibfield  {journal} {\bibinfo  {journal}
  {Proc. Natl. Acad. Sci. U. S. A.}\ }\textbf {\bibinfo {volume} {108}},\
  \bibinfo {pages} {11772} (\bibinfo {year} {2011})}\BibitemShut {NoStop}%
\bibitem [{\citenamefont {Misbah}(2012)}]{misb12}%
  \BibitemOpen
  \bibfield  {author} {\bibinfo {author} {\bibfnamefont {C.}~\bibnamefont
  {Misbah}},\ }\href {\doibase 10.1088/1742-6596/392/1/012005} {\bibfield
  {journal} {\bibinfo  {journal} {J. Phys.: Conf. Ser.}\ }\textbf {\bibinfo
  {volume} {392}},\ \bibinfo {pages} {012005} (\bibinfo {year}
  {2012})}\BibitemShut {NoStop}%
\bibitem [{\citenamefont {Dupire}\ \emph {et~al.}(2012)\citenamefont {Dupire},
  \citenamefont {Socol},\ and\ \citenamefont {Viallat}}]{dupi12}%
  \BibitemOpen
  \bibfield  {author} {\bibinfo {author} {\bibfnamefont {J.}~\bibnamefont
  {Dupire}}, \bibinfo {author} {\bibfnamefont {M.}~\bibnamefont {Socol}}, \
  and\ \bibinfo {author} {\bibfnamefont {A.}~\bibnamefont {Viallat}},\ }\href
  {\doibase 10.1073/pnas.1210236109} {\bibfield  {journal} {\bibinfo  {journal}
  {Proc. Nat. Acad. Sci. U. S. A.}\ }\textbf {\bibinfo {volume} {109}},\
  \bibinfo {pages} {20808–20813} (\bibinfo {year} {2012})}\BibitemShut
  {NoStop}%
\bibitem [{\citenamefont {Forsyth}\ \emph {et~al.}(2011)\citenamefont
  {Forsyth}, \citenamefont {Wan}, \citenamefont {Owrutsky}, \citenamefont
  {Abkarian},\ and\ \citenamefont {Stone}}]{fors11}%
  \BibitemOpen
  \bibfield  {author} {\bibinfo {author} {\bibfnamefont {A.~M.}\ \bibnamefont
  {Forsyth}}, \bibinfo {author} {\bibfnamefont {J.}~\bibnamefont {Wan}},
  \bibinfo {author} {\bibfnamefont {P.~D.}\ \bibnamefont {Owrutsky}}, \bibinfo
  {author} {\bibfnamefont {M.}~\bibnamefont {Abkarian}}, \ and\ \bibinfo
  {author} {\bibfnamefont {H.~A.}\ \bibnamefont {Stone}},\ }\href {\doibase
  10.1073/pnas.1101315108} {\bibfield  {journal} {\bibinfo  {journal} {Proc.
  Natl. Acad. Sci. U. S. A.}\ }\textbf {\bibinfo {volume} {108}},\ \bibinfo
  {pages} {10986} (\bibinfo {year} {2011})}\BibitemShut {NoStop}%
\bibitem [{\citenamefont {Vlahovska}\ and\ \citenamefont
  {Gracia}(2007)}]{vlah07}%
  \BibitemOpen
  \bibfield  {author} {\bibinfo {author} {\bibfnamefont {P.~M.}\ \bibnamefont
  {Vlahovska}}\ and\ \bibinfo {author} {\bibfnamefont {R.~S.}\ \bibnamefont
  {Gracia}},\ }\href {\doibase 10.1103/PhysRevE.75.016313} {\bibfield
  {journal} {\bibinfo  {journal} {Phys. Rev. E}\ }\textbf {\bibinfo {volume}
  {75}},\ \bibinfo {pages} {016313} (\bibinfo {year} {2007})}\BibitemShut
  {NoStop}%
\bibitem [{\citenamefont {Lebedev}\ \emph {et~al.}(2007)\citenamefont
  {Lebedev}, \citenamefont {Turitsyn},\ and\ \citenamefont
  {Vergeles}}]{lebe07}%
  \BibitemOpen
  \bibfield  {author} {\bibinfo {author} {\bibfnamefont {V.~V.}\ \bibnamefont
  {Lebedev}}, \bibinfo {author} {\bibfnamefont {K.~S.}\ \bibnamefont
  {Turitsyn}}, \ and\ \bibinfo {author} {\bibfnamefont {S.~S.}\ \bibnamefont
  {Vergeles}},\ }\href {\doibase 10.1103/PhysRevLett.99.218101} {\bibfield
  {journal} {\bibinfo  {journal} {Phys. Rev. Lett.}\ }\textbf {\bibinfo
  {volume} {99}},\ \bibinfo {pages} {218101} (\bibinfo {year}
  {2007})}\BibitemShut {NoStop}%
\bibitem [{\citenamefont {Lebedev}\ \emph {et~al.}(2008)\citenamefont
  {Lebedev}, \citenamefont {Turitsyn},\ and\ \citenamefont
  {Vergeles}}]{lebe08}%
  \BibitemOpen
  \bibfield  {author} {\bibinfo {author} {\bibfnamefont {V.~V.}\ \bibnamefont
  {Lebedev}}, \bibinfo {author} {\bibfnamefont {K.~S.}\ \bibnamefont
  {Turitsyn}}, \ and\ \bibinfo {author} {\bibfnamefont {S.~S.}\ \bibnamefont
  {Vergeles}},\ }\href {\doibase 10.1088/1367-2630/10/4/043044} {\bibfield
  {journal} {\bibinfo  {journal} {New J. Phys.}\ }\textbf {\bibinfo {volume}
  {10}},\ \bibinfo {pages} {043044} (\bibinfo {year} {2008})}\BibitemShut
  {NoStop}%
\bibitem [{\citenamefont {Farutin}\ \emph {et~al.}(2010)\citenamefont
  {Farutin}, \citenamefont {Biben},\ and\ \citenamefont {Misbah}}]{faru10}%
  \BibitemOpen
  \bibfield  {author} {\bibinfo {author} {\bibfnamefont {A.}~\bibnamefont
  {Farutin}}, \bibinfo {author} {\bibfnamefont {T.}~\bibnamefont {Biben}}, \
  and\ \bibinfo {author} {\bibfnamefont {C.}~\bibnamefont {Misbah}},\ }\href
  {\doibase 10.1103/PhysRevE.81.061904} {\bibfield  {journal} {\bibinfo
  {journal} {Phys. Rev. E}\ }\textbf {\bibinfo {volume} {81}},\ \bibinfo
  {pages} {061904} (\bibinfo {year} {2010})}\BibitemShut {NoStop}%
\bibitem [{\citenamefont {Farutin}\ \emph {et~al.}(2012)\citenamefont
  {Farutin}, \citenamefont {Aouane},\ and\ \citenamefont {Misbah}}]{faru12a}%
  \BibitemOpen
  \bibfield  {author} {\bibinfo {author} {\bibfnamefont {A.}~\bibnamefont
  {Farutin}}, \bibinfo {author} {\bibfnamefont {O.}~\bibnamefont {Aouane}}, \
  and\ \bibinfo {author} {\bibfnamefont {C.}~\bibnamefont {Misbah}},\ }\href
  {\doibase 10.1103/PhysRevE.85.061922} {\bibfield  {journal} {\bibinfo
  {journal} {Phys. Rev. E}\ }\textbf {\bibinfo {volume} {85}},\ \bibinfo
  {pages} {061922} (\bibinfo {year} {2012})}\BibitemShut {NoStop}%
\bibitem [{\citenamefont {Farutin}\ and\ \citenamefont
  {Misbah}(2012)}]{faru12b}%
  \BibitemOpen
  \bibfield  {author} {\bibinfo {author} {\bibfnamefont {A.}~\bibnamefont
  {Farutin}}\ and\ \bibinfo {author} {\bibfnamefont {C.}~\bibnamefont
  {Misbah}},\ }\href {\doibase 10.1103/PhysRevLett.109.248106} {\bibfield
  {journal} {\bibinfo  {journal} {Phys. Rev. Lett.}\ }\textbf {\bibinfo
  {volume} {109}},\ \bibinfo {pages} {248106} (\bibinfo {year}
  {2012})}\BibitemShut {NoStop}%
\bibitem [{\citenamefont {Biben}\ \emph {et~al.}(2011)\citenamefont {Biben},
  \citenamefont {Farutin},\ and\ \citenamefont {Misbah}}]{bibe11}%
  \BibitemOpen
  \bibfield  {author} {\bibinfo {author} {\bibfnamefont {T.}~\bibnamefont
  {Biben}}, \bibinfo {author} {\bibfnamefont {A.}~\bibnamefont {Farutin}}, \
  and\ \bibinfo {author} {\bibfnamefont {C.}~\bibnamefont {Misbah}},\ }\href
  {\doibase 10.1103/PhysRevE.83.031921} {\bibfield  {journal} {\bibinfo
  {journal} {Phys. Rev. E}\ }\textbf {\bibinfo {volume} {83}},\ \bibinfo
  {pages} {031921} (\bibinfo {year} {2011})}\BibitemShut {NoStop}%
\bibitem [{\citenamefont {Zhao}\ and\ \citenamefont {Shaqfeh}(2011)}]{zhao11}%
  \BibitemOpen
  \bibfield  {author} {\bibinfo {author} {\bibfnamefont {H.}~\bibnamefont
  {Zhao}}\ and\ \bibinfo {author} {\bibfnamefont {E.~S.~G.}\ \bibnamefont
  {Shaqfeh}},\ }\href {\doibase 10.1017/S0022112011000115} {\bibfield
  {journal} {\bibinfo  {journal} {J. Fluid. Mech.}\ }\textbf {\bibinfo {volume}
  {674}},\ \bibinfo {pages} {578} (\bibinfo {year} {2011})}\BibitemShut
  {NoStop}%
\bibitem [{\citenamefont {Salac}\ and\ \citenamefont {Miksis}(2012)}]{sala12}%
  \BibitemOpen
  \bibfield  {author} {\bibinfo {author} {\bibfnamefont {D.}~\bibnamefont
  {Salac}}\ and\ \bibinfo {author} {\bibfnamefont {M.~J.}\ \bibnamefont
  {Miksis}},\ }\href {\doibase 10.1017/jfm.2012.380} {\bibfield  {journal}
  {\bibinfo  {journal} {J. Fluid Mech.}\ }\textbf {\bibinfo {volume} {711}},\
  \bibinfo {pages} {122} (\bibinfo {year} {2012})}\BibitemShut {NoStop}%
\bibitem [{\citenamefont {Yazdani}\ and\ \citenamefont
  {Bagchi}(2012)}]{yazd12}%
  \BibitemOpen
  \bibfield  {author} {\bibinfo {author} {\bibfnamefont {A.}~\bibnamefont
  {Yazdani}}\ and\ \bibinfo {author} {\bibfnamefont {P.}~\bibnamefont
  {Bagchi}},\ }\href {\doibase 10.1103/PhysRevE.85.056308} {\bibfield
  {journal} {\bibinfo  {journal} {Phys. Rev. E}\ }\textbf {\bibinfo {volume}
  {85}},\ \bibinfo {pages} {056308} (\bibinfo {year} {2012})}\BibitemShut
  {NoStop}%
\bibitem [{\citenamefont {Misbah}(2006)}]{misb06}%
  \BibitemOpen
  \bibfield  {author} {\bibinfo {author} {\bibfnamefont {C.}~\bibnamefont
  {Misbah}},\ }\href {\doibase 10.1103/PhysRevLett.96.028104} {\bibfield
  {journal} {\bibinfo  {journal} {Phys. Rev. Lett.}\ }\textbf {\bibinfo
  {volume} {96}},\ \bibinfo {pages} {028104} (\bibinfo {year}
  {2006})}\BibitemShut {NoStop}%
\bibitem [{\citenamefont {Guedda}\ \emph {et~al.}(2012)\citenamefont {Guedda},
  \citenamefont {Abaidi}, \citenamefont {Benlahsen},\ and\ \citenamefont
  {Misbah}}]{gued12}%
  \BibitemOpen
  \bibfield  {author} {\bibinfo {author} {\bibfnamefont {M.}~\bibnamefont
  {Guedda}}, \bibinfo {author} {\bibfnamefont {M.}~\bibnamefont {Abaidi}},
  \bibinfo {author} {\bibfnamefont {M.}~\bibnamefont {Benlahsen}}, \ and\
  \bibinfo {author} {\bibfnamefont {C.}~\bibnamefont {Misbah}},\ }\href
  {\doibase 10.1103/PhysRevE.86.051915} {\bibfield  {journal} {\bibinfo
  {journal} {Phys. Rev. E}\ }\textbf {\bibinfo {volume} {86}},\ \bibinfo
  {pages} {051915} (\bibinfo {year} {2012})}\BibitemShut {NoStop}%
\bibitem [{\citenamefont {Noguchi}\ and\ \citenamefont
  {Gompper}(2007)}]{nogu07}%
  \BibitemOpen
  \bibfield  {author} {\bibinfo {author} {\bibfnamefont {H.}~\bibnamefont
  {Noguchi}}\ and\ \bibinfo {author} {\bibfnamefont {G.}~\bibnamefont
  {Gompper}},\ }\href {\doibase 10.1103/PhysRevLett.98.128103} {\bibfield
  {journal} {\bibinfo  {journal} {Phys. Rev. Lett.}\ }\textbf {\bibinfo
  {volume} {98}},\ \bibinfo {pages} {128103} (\bibinfo {year}
  {2007})}\BibitemShut {NoStop}%
\bibitem [{\citenamefont {Kantsler}\ and\ \citenamefont
  {Steinberg}(2005)}]{kant05}%
  \BibitemOpen
  \bibfield  {author} {\bibinfo {author} {\bibfnamefont {V.}~\bibnamefont
  {Kantsler}}\ and\ \bibinfo {author} {\bibfnamefont {V.}~\bibnamefont
  {Steinberg}},\ }\href {\doibase 10.1103/PhysRevLett.95.258101} {\bibfield
  {journal} {\bibinfo  {journal} {Phys. Rev. Lett.}\ }\textbf {\bibinfo
  {volume} {95}},\ \bibinfo {pages} {258101} (\bibinfo {year}
  {2005})}\BibitemShut {NoStop}%
\bibitem [{\citenamefont {Kantsler}\ and\ \citenamefont
  {Steinberg}(2006)}]{kant06}%
  \BibitemOpen
  \bibfield  {author} {\bibinfo {author} {\bibfnamefont {V.}~\bibnamefont
  {Kantsler}}\ and\ \bibinfo {author} {\bibfnamefont {V.}~\bibnamefont
  {Steinberg}},\ }\href {\doibase 10.1103/PhysRevLett.96.036001} {\bibfield
  {journal} {\bibinfo  {journal} {Phys. Rev. Lett.}\ }\textbf {\bibinfo
  {volume} {96}},\ \bibinfo {pages} {036001} (\bibinfo {year}
  {2006})}\BibitemShut {NoStop}%
\bibitem [{\citenamefont {Mader}\ \emph {et~al.}(2006)\citenamefont {Mader},
  \citenamefont {Vitkova}, \citenamefont {Abkarian}, \citenamefont {Viallat},\
  and\ \citenamefont {Podgorski}}]{made06}%
  \BibitemOpen
  \bibfield  {author} {\bibinfo {author} {\bibfnamefont {M.}~\bibnamefont
  {Mader}}, \bibinfo {author} {\bibfnamefont {V.}~\bibnamefont {Vitkova}},
  \bibinfo {author} {\bibfnamefont {M.}~\bibnamefont {Abkarian}}, \bibinfo
  {author} {\bibfnamefont {A.}~\bibnamefont {Viallat}}, \ and\ \bibinfo
  {author} {\bibfnamefont {T.}~\bibnamefont {Podgorski}},\ }\href {\doibase
  10.1140/epje/i2005-10058-x} {\bibfield  {journal} {\bibinfo  {journal} {Eur.
  Phys. J. E}\ }\textbf {\bibinfo {volume} {19}},\ \bibinfo {pages} {389}
  (\bibinfo {year} {2006})}\BibitemShut {NoStop}%
\bibitem [{\citenamefont {Deschamps}\ \emph
  {et~al.}(2009{\natexlab{a}})\citenamefont {Deschamps}, \citenamefont
  {Kantsler},\ and\ \citenamefont {Steinberg}}]{desc09}%
  \BibitemOpen
  \bibfield  {author} {\bibinfo {author} {\bibfnamefont {J.}~\bibnamefont
  {Deschamps}}, \bibinfo {author} {\bibfnamefont {V.}~\bibnamefont {Kantsler}},
  \ and\ \bibinfo {author} {\bibfnamefont {V.}~\bibnamefont {Steinberg}},\
  }\href {\doibase 10.1103/PhysRevLett.102.118105} {\bibfield  {journal}
  {\bibinfo  {journal} {Phys. Rev. Lett.}\ }\textbf {\bibinfo {volume} {102}},\
  \bibinfo {pages} {118105} (\bibinfo {year} {2009}{\natexlab{a}})}\BibitemShut
  {NoStop}%
\bibitem [{\citenamefont {Deschamps}\ \emph
  {et~al.}(2009{\natexlab{b}})\citenamefont {Deschamps}, \citenamefont
  {Kantsler}, \citenamefont {Segre},\ and\ \citenamefont
  {Steinberg}}]{desc09a}%
  \BibitemOpen
  \bibfield  {author} {\bibinfo {author} {\bibfnamefont {J.}~\bibnamefont
  {Deschamps}}, \bibinfo {author} {\bibfnamefont {V.}~\bibnamefont {Kantsler}},
  \bibinfo {author} {\bibfnamefont {E.}~\bibnamefont {Segre}}, \ and\ \bibinfo
  {author} {\bibfnamefont {V.}~\bibnamefont {Steinberg}},\ }\href {\doibase
  10.1073/pnas.0902657106} {\bibfield  {journal} {\bibinfo  {journal} {Proc.
  Natl. Acad. Sci. U. S. A.}\ }\textbf {\bibinfo {volume} {106}},\ \bibinfo
  {pages} {11444} (\bibinfo {year} {2009}{\natexlab{b}})}\BibitemShut {NoStop}%
\bibitem [{\citenamefont {Zabusky}\ \emph {et~al.}(2011)\citenamefont
  {Zabusky}, \citenamefont {Segre}, \citenamefont {Deschamps}, \citenamefont
  {Kantsler},\ and\ \citenamefont {Steinberg}}]{zabu11}%
  \BibitemOpen
  \bibfield  {author} {\bibinfo {author} {\bibfnamefont {N.~J.}\ \bibnamefont
  {Zabusky}}, \bibinfo {author} {\bibfnamefont {E.}~\bibnamefont {Segre}},
  \bibinfo {author} {\bibfnamefont {J.}~\bibnamefont {Deschamps}}, \bibinfo
  {author} {\bibfnamefont {V.}~\bibnamefont {Kantsler}}, \ and\ \bibinfo
  {author} {\bibfnamefont {V.}~\bibnamefont {Steinberg}},\ }\href {\doibase
  10.1063/1.3556439} {\bibfield  {journal} {\bibinfo  {journal} {Phys. Fluids}\
  }\textbf {\bibinfo {volume} {23}},\ \bibinfo {eid} {041905} (\bibinfo {year}
  {2011})}\BibitemShut {NoStop}%
\bibitem [{\citenamefont {Noguchi}\ and\ \citenamefont
  {Gompper}(2004)}]{nogu04}%
  \BibitemOpen
  \bibfield  {author} {\bibinfo {author} {\bibfnamefont {H.}~\bibnamefont
  {Noguchi}}\ and\ \bibinfo {author} {\bibfnamefont {G.}~\bibnamefont
  {Gompper}},\ }\href {\doibase 10.1103/PhysRevLett.93.258102} {\bibfield
  {journal} {\bibinfo  {journal} {Phys. Rev. Lett.}\ }\textbf {\bibinfo
  {volume} {93}},\ \bibinfo {pages} {258102} (\bibinfo {year}
  {2004})}\BibitemShut {NoStop}%
\bibitem [{\citenamefont {Noguchi}\ and\ \citenamefont
  {Gompper}(2005)}]{nogu05}%
  \BibitemOpen
  \bibfield  {author} {\bibinfo {author} {\bibfnamefont {H.}~\bibnamefont
  {Noguchi}}\ and\ \bibinfo {author} {\bibfnamefont {G.}~\bibnamefont
  {Gompper}},\ }\href {\doibase 10.1103/PhysRevE.72.011901} {\bibfield
  {journal} {\bibinfo  {journal} {Phys. Rev. E}\ }\textbf {\bibinfo {volume}
  {72}},\ \bibinfo {pages} {011901} (\bibinfo {year} {2005})}\BibitemShut
  {NoStop}%
\bibitem [{\citenamefont {Me{\ss}linger}\ \emph {et~al.}(2009)\citenamefont
  {Me{\ss}linger}, \citenamefont {Schmidt}, \citenamefont {Noguchi},\ and\
  \citenamefont {Gompper}}]{me09}%
  \BibitemOpen
  \bibfield  {author} {\bibinfo {author} {\bibfnamefont {S.}~\bibnamefont
  {Me{\ss}linger}}, \bibinfo {author} {\bibfnamefont {B.}~\bibnamefont
  {Schmidt}}, \bibinfo {author} {\bibfnamefont {H.}~\bibnamefont {Noguchi}}, \
  and\ \bibinfo {author} {\bibfnamefont {G.}~\bibnamefont {Gompper}},\ }\href
  {\doibase 10.1103/PhysRevE.80.011901} {\bibfield  {journal} {\bibinfo
  {journal} {Phys. Rev. E}\ }\textbf {\bibinfo {volume} {80}},\ \bibinfo
  {pages} {011901} (\bibinfo {year} {2009})}\BibitemShut {NoStop}%
\bibitem [{\citenamefont {Seifert}(1999)}]{seif99}%
  \BibitemOpen
  \bibfield  {author} {\bibinfo {author} {\bibfnamefont {U.}~\bibnamefont
  {Seifert}},\ }\href {\doibase 10.1007/s100510050706} {\bibfield  {journal}
  {\bibinfo  {journal} {Eur. Phys. J. B}\ }\textbf {\bibinfo {volume} {8}},\
  \bibinfo {pages} {405} (\bibinfo {year} {1999})}\BibitemShut {NoStop}%
\bibitem [{\citenamefont {Finken}\ \emph {et~al.}(2008)\citenamefont {Finken},
  \citenamefont {Lamura}, \citenamefont {Seifert},\ and\ \citenamefont
  {Gompper}}]{fink08}%
  \BibitemOpen
  \bibfield  {author} {\bibinfo {author} {\bibfnamefont {R.}~\bibnamefont
  {Finken}}, \bibinfo {author} {\bibfnamefont {A.}~\bibnamefont {Lamura}},
  \bibinfo {author} {\bibfnamefont {U.}~\bibnamefont {Seifert}}, \ and\
  \bibinfo {author} {\bibfnamefont {G.}~\bibnamefont {Gompper}},\ }\href
  {\doibase 10.1140/epje/i2007-10299-7} {\bibfield  {journal} {\bibinfo
  {journal} {Eur. Phys. J. E}\ }\textbf {\bibinfo {volume} {25}},\ \bibinfo
  {pages} {309} (\bibinfo {year} {2008})}\BibitemShut {NoStop}%
\bibitem [{\citenamefont {Abreu}\ and\ \citenamefont {Seifert}(2012)}]{abre12}%
  \BibitemOpen
  \bibfield  {author} {\bibinfo {author} {\bibfnamefont {D.}~\bibnamefont
  {Abreu}}\ and\ \bibinfo {author} {\bibfnamefont {U.}~\bibnamefont
  {Seifert}},\ }\href {\doibase 10.1103/PhysRevE.86.010902} {\bibfield
  {journal} {\bibinfo  {journal} {Phys. Rev. E}\ }\textbf {\bibinfo {volume}
  {86}},\ \bibinfo {pages} {010902} (\bibinfo {year} {2012})}\BibitemShut
  {NoStop}%
\bibitem [{\citenamefont {Zhong-{C}an}\ and\ \citenamefont
  {Helfrich}(1989)}]{zho89}%
  \BibitemOpen
  \bibfield  {author} {\bibinfo {author} {\bibfnamefont {O.-Y.}\ \bibnamefont
  {Zhong-{C}an}}\ and\ \bibinfo {author} {\bibfnamefont {W.}~\bibnamefont
  {Helfrich}},\ }\href {\doibase 10.1103/PhysRevA.39.5280} {\bibfield
  {journal} {\bibinfo  {journal} {Phys. Rev. A}\ }\textbf {\bibinfo {volume}
  {39}},\ \bibinfo {pages} {5280} (\bibinfo {year} {1989})}\BibitemShut
  {NoStop}%
\bibitem [{\citenamefont {Danker}\ \emph {et~al.}(2007)\citenamefont {Danker},
  \citenamefont {Biben}, \citenamefont {Podgorski}, \citenamefont {Verdier},\
  and\ \citenamefont {Misbah}}]{dank07}%
  \BibitemOpen
  \bibfield  {author} {\bibinfo {author} {\bibfnamefont {G.}~\bibnamefont
  {Danker}}, \bibinfo {author} {\bibfnamefont {T.}~\bibnamefont {Biben}},
  \bibinfo {author} {\bibfnamefont {T.}~\bibnamefont {Podgorski}}, \bibinfo
  {author} {\bibfnamefont {C.}~\bibnamefont {Verdier}}, \ and\ \bibinfo
  {author} {\bibfnamefont {C.}~\bibnamefont {Misbah}},\ }\href {\doibase
  10.1103/PhysRevE.76.041905} {\bibfield  {journal} {\bibinfo  {journal} {Phys.
  Rev. E}\ }\textbf {\bibinfo {volume} {76}},\ \bibinfo {pages} {041905}
  (\bibinfo {year} {2007})}\BibitemShut {NoStop}%
\bibitem [{\citenamefont {Morse}(2004)}]{mors04}%
  \BibitemOpen
  \bibfield  {author} {\bibinfo {author} {\bibfnamefont {D.~C.}\ \bibnamefont
  {Morse}},\ }\href@noop {} {\bibfield  {journal} {\bibinfo  {journal} {Adv.
  Chem. Phys.}\ }\textbf {\bibinfo {volume} {128}},\ \bibinfo {pages} {65}
  (\bibinfo {year} {2004})}\BibitemShut {NoStop}%
\bibitem [{\citenamefont {Kaoui}\ \emph {et~al.}(2009)\citenamefont {Kaoui},
  \citenamefont {Farutin},\ and\ \citenamefont {Misbah}}]{kaou09a}%
  \BibitemOpen
  \bibfield  {author} {\bibinfo {author} {\bibfnamefont {B.}~\bibnamefont
  {Kaoui}}, \bibinfo {author} {\bibfnamefont {A.}~\bibnamefont {Farutin}}, \
  and\ \bibinfo {author} {\bibfnamefont {C.}~\bibnamefont {Misbah}},\ }\href
  {\doibase 10.1103/PhysRevE.80.061905} {\bibfield  {journal} {\bibinfo
  {journal} {Phys. Rev. E}\ }\textbf {\bibinfo {volume} {80}},\ \bibinfo
  {pages} {061905} (\bibinfo {year} {2009})}\BibitemShut {NoStop}%
\bibitem [{\citenamefont {Kantsler}\ \emph {et~al.}(2007)\citenamefont
  {Kantsler}, \citenamefont {Segre},\ and\ \citenamefont {Steinberg}}]{kant07}%
  \BibitemOpen
  \bibfield  {author} {\bibinfo {author} {\bibfnamefont {V.}~\bibnamefont
  {Kantsler}}, \bibinfo {author} {\bibfnamefont {E.}~\bibnamefont {Segre}}, \
  and\ \bibinfo {author} {\bibfnamefont {V.}~\bibnamefont {Steinberg}},\ }\href
  {\doibase 10.1103/PhysRevLett.99.178102} {\bibfield  {journal} {\bibinfo
  {journal} {Phys. Rev. Lett.}\ }\textbf {\bibinfo {volume} {99}},\ \bibinfo
  {pages} {178102} (\bibinfo {year} {2007})}\BibitemShut {NoStop}%
\bibitem [{\citenamefont {Finken}\ and\ \citenamefont
  {Seifert}(2006)}]{fink06}%
  \BibitemOpen
  \bibfield  {author} {\bibinfo {author} {\bibfnamefont {R.}~\bibnamefont
  {Finken}}\ and\ \bibinfo {author} {\bibfnamefont {U.}~\bibnamefont
  {Seifert}},\ }\href {\doibase 10.1088/0953-8984/18/15/L04} {\bibfield
  {journal} {\bibinfo  {journal} {J. Phys.: Condens. Matter}\ }\textbf
  {\bibinfo {volume} {18}},\ \bibinfo {pages} {L185} (\bibinfo {year}
  {2006})}\BibitemShut {NoStop}%
\bibitem [{\citenamefont {Turitsyn}\ and\ \citenamefont
  {Vergeles}(2008)}]{turi08}%
  \BibitemOpen
  \bibfield  {author} {\bibinfo {author} {\bibfnamefont {K.~S.}\ \bibnamefont
  {Turitsyn}}\ and\ \bibinfo {author} {\bibfnamefont {S.~S.}\ \bibnamefont
  {Vergeles}},\ }\href {\doibase 10.1103/PhysRevLett.100.028103} {\bibfield
  {journal} {\bibinfo  {journal} {Phys. Rev. Lett.}\ }\textbf {\bibinfo
  {volume} {100}},\ \bibinfo {pages} {028103} (\bibinfo {year}
  {2008})}\BibitemShut {NoStop}%
\end{thebibliography}%

\end{document}